\documentclass[prb,aps]{revtex4}

\usepackage[dvips]{graphics}
\usepackage{amssymb}
\usepackage{psfig}

\begin{document}

\draft

\title{Coherent electronic transport through a superconducting film}
\author{M. Bo\v{z}ovi\'{c}, Z. Pajovi\'c, and Z. Radovi\'c}
\address{Department of Physics, University of Belgrade, P.O. Box 368, 11001 Belgrade, Yugoslavia}

\begin{abstract}
We study coherent quantum transport through a superconducting film
connected to normal-metal electrodes. Simple expressions for the
differential conductance and the local density of states are
obtained in the clean limit and for transparent interfaces.
Quasiparticle interference causes periodic vanishing of the
Andreev reflection at the energies of geometrical resonances,
subgap transport, and gapless superconductivity near the
interfaces. Application of the results to spectroscopic
measurements of the superconducting gap and the Fermi velocity is
analyzed.
\end{abstract}


\maketitle

\section{Introduction}

Conducting properties of normal metal/superconductor/normal metal
(NSN) heterostructures have attracted considerable
interest.\cite{Beenakker,NBL} Electronic transport through a
normal metal/superconductor (NS) junction, with an insulating
barrier of arbitrary strength at the interface, was studied by
Blonder, Tinkham, and Klapwijk (BTK),\cite{BTK} and the Andreev
reflection\cite{Andreev} was recognized as the mechanism of
normal-to-supercurrent conversion.\cite{Ishii,Furusaki Tsukada}
The BTK model can be used when electrons pass incoherently through
the interfaces of NSN double junctions.\cite{Zheng}

However, in clean superconducting heterostructures coherent
quantum transport is strongly influenced by size effects due to
the interplay between geometrical resonances and the Andreev
reflection.\cite{Sanchez,Brinkman,Milos} Well-known examples are
the current-carrying Andreev bound states\cite{Sipr,Tanaka 00} and
multiple Andreev reflections\cite{KBT,Basel,Ingerman} in
superconductor/normal metal/superconductor (SNS) junctions. Since
early experiments by Tomasch,\cite{Tomasch} the geometric
resonance nature of the differential conductance oscillations in
SNS and NSN tunnel junctions has been attributed to the electron
interference in the central
layer.\cite{Anderson,Rowell,McMillan,Demers} Recently,
McMillan-Rowell oscillations were observed in SNS tunnel junctions
of $d$-wave superconductors, and used in measurements of the
superconducting gap and the Fermi velocity.\cite{Nesher}

In this paper, we focus on size and coherency effects in NSN
double junctions with a clean conventional ($s$-wave)
superconductor and transparent interfaces. In that case, instead
of a fully numerical treatment,\cite{Sanchez,Stojkovic} a simple
analytic form for the differential conductance and the local
quasiparticle density of states (DOS) can be obtained from the
solution of the Bogoliubov-de Gennes equation. The proximity
effect, which is significant in thin superconducting films, is
taken into account through the self-consistent calculation of the
pair potential. Striking consequence of the quasiparticle
interference in the clean metallic NSN junctions include periodic
vanishing of the Andreev reflection at the energies of geometrical
resonances and the characteristic changes of DOS with respect to
the bulk superconductor.

\section{Scattering problem}

We consider an NSN double junction consisting of a clean
superconducting layer of thickness $d$, connected to normal-metal
electrodes by transparent interfaces. The quasiparticle
propagation is described by the Bogoliubov--de Gennes equation
\begin{eqnarray}
\left(
\begin{array}{ccc}
  H_0({\bf r}) && \Delta({\bf r}) \\
  \Delta^{*}({\bf r}) && -H_0({\bf r})
\end{array}
\right) \left(\begin{array}{c}
    u({\bf r}) \\
    v({\bf r})
  \end{array}\right)~=~E\left(\begin{array}{c}
    u({\bf r}) \\
    v({\bf r})
  \end{array}\right), \label{BdG}
\end{eqnarray}
with $H_{0}({\bf r})=-\hbar^{2}\nabla^{2}/2m-E_F$, where $E_F$ is
the Fermi energy. For simplicity, the magnitude of the Fermi wave
vector, $k_F=\sqrt{2mE_F/\hbar^2}$, is assumed to be constant
through the junction. The superconducting pair potential is taken
in the form $\Delta({\bf r})=\Delta \Theta(z) \Theta(l-z)$, where
$\Theta(z)$ is the Heaviside step function, and the $z$-axis is
perpendicular to the layers. In Eq. (\ref{BdG}), $E$ is the
quasiparticle energy with respect to the Fermi level. The electron
effective mass $m$ is assumed to be the same for the entire
junction. The parallel component of the wave vector ${\bf k}_{||}$
is conserved, and the wave function
\begin{equation}
\left(\begin{array}{c}
    u({\bf r}) \\
    v({\bf r})
  \end{array}\right)=\exp(i{\bf k}_{||} \cdot {\bf r}
  )~\psi (z),
\end{equation}
satisfies the continuity boundary conditions for $\psi (z)$ and
$\psi '(z)$ at $z=0$ and $z=d$. The four independent solutions of
Eq. (\ref{BdG}) correspond to the four types of injection: an
electron or a hole, from either the left or the right
electrode.\cite{Furusaki Tsukada}

For the injection of an electron from the left, with energy $E>0$
and angle of incidence $\theta$ (measured from the $z$-axis),
solution for $\psi (z)$ has the following form
\begin{eqnarray}
\label{psiL} \psi (z)=
  \left\{
  \begin{array}{ll}
     \exp(ik^+z)\left(\begin{array}{c}
     1 \\
     0 \\
   \end{array}\right)+
    a(E,\theta)\exp(ik^-z)\left(\begin{array}{c}
      0 \\
      1 \\
    \end{array}\right)  & z < 0,   \\
    \\
b_{1}(E,\theta)\exp(iq^+z) \left(\begin{array}{c}
    \bar{u} \\
    \bar{v}
  \end{array}\right)+b_{2}(E,\theta)\exp(iq^-z)
\left(\begin{array}{c}
    \bar{v} \\
    \bar{u}
  \end{array}\right)  & 0<z<d,  \\
  \\
 c(E,\theta)\exp(ik^+z)\left(\begin{array}{c}
     1 \\
     0 \\
   \end{array}\right) & z>d.
   \\
   \end{array}
   \right.
\end{eqnarray}
Here, $\bar{u}=\sqrt{(1+\Omega/E)/2}$ and
$\bar{v}=\sqrt{(1-\Omega/E)/2}$ are the BCS coherence factors, and
$\Omega=\sqrt{E^2-\Delta^2}$. The $z$-components of the wave
vectors are $k^\pm=\sqrt{(2m/\hbar ^2)(E_F\pm E)-{\bf k}^2_{||}}$
and $q^\pm=\sqrt{(2m/\hbar ^2)(E_F\pm\Omega)-{\bf k}^2_{||}}$,
where $|{\bf k}_{||}|=\sqrt{(2m/\hbar ^2)(E_F+E)}~\sin\theta$. The
coefficients $a$ and $c$ are, respectively, the probability
amplitudes of the Andreev reflection as a hole of the opposite
spin (AR) and transmission to the right electrode as an electron
(TE). The amplitudes of the Bogoliubov electron-like and hole-like
quasiparticles, propagating in the superconducting layer, are
given by the coefficients $b_1$ and $b_2$. The normal reflection
of electrons and the transmission to the right electrode as a hole
are absent. Solutions for the general case of insulating
interfaces and the Fermi velocity mismatch are given in Ref. 7.

By neglecting the small terms $E/E_F\ll 1$ and $\Delta/E_F\ll 1$
in the wave vectors, except in the exponent
\begin{equation}
\label{zeta} \zeta =d\left(q^+ - q^-\right),
\end{equation}
solutions of the boundary-condition equations can be written in a
simple form: $a(E,\theta)=\bar{u}\bar{v}[1-\exp(i\zeta
z/d)]/[\bar{u}^2-\bar{v}^2\exp(i\zeta z/d)]$,
$b_1(E,\theta)=\bar{u}/[\bar{u}^2-\bar{v}^2\exp(i\zeta z/d)]$,
$b_2(E,\theta)=-\bar{v}\exp(i\zeta
z/d)/[\bar{u}^2-\bar{v}^2\exp(i\zeta z/d)]$, and
$c(E,\theta)=-i(\bar{u}^2-\bar{v}^2)\exp(i\zeta
z/d)/[\bar{u}^2-\bar{v}^2\exp(i\zeta z/d)]$. From the probability
current conservation, the probabilities of outgoing particles
satisfy the normalization condition $A(E,\theta)+C(E,\theta)=1$,
where $A(E,\theta)=|a(E,\theta)|^2$ and
$C(E,\theta)=|c(E,\theta)|^2$. Explicitly, the AR probability is
\begin{equation}
\label{a}
A(E,\theta)=\left|\frac{\Delta\sin(\zeta/2)}{E\sin(\zeta/2)+i\Omega\cos(\zeta/2)}\right|^2.
\end{equation}

Solutions for the other three types of injection can be obtained
by the same procedure. In particular, if a hole with energy $-E$
and angle of incidence $\theta$ is injected from the left, the
substitution $q^+ \rightleftharpoons q^-$ holds, and the
scattering probabilities are the same as for the injection of an
electron with $E$ and $\theta$. Therefore, in order to include the
description of both electron and hole injection, the calculated
probabilities should be regarded as even functions of $E$. Also,
for an electron or a hole injected from the right, the
probabilities are the same as for the injection from the left.

From Eq. (\ref{a}), it follows that $A(E,\theta)=0$, and
consequently $C(E,\theta)=1$, when
\begin{equation}
\label{resonance} \zeta=2n\pi
\end{equation}
for $n=0,\pm 1,\pm 2,\ldots$. Therefore, the Andreev reflection
vanishes, and the electron transmission becomes complete (without
creation or annihilation of Cooper pairs), at the energies of
geometrical resonances in the quasiparticle spectrum, given by Eq.
(\ref{resonance}).

For thin superconducting films, the proximity effect is important
and the self-consistent treatment of the pair potential is
required. However, the spatial variation of the pair potential is
negligible for $d/\xi_0\lesssim 1$, where $\xi_0=\hbar
v_F/\pi\Delta_0$ is the BCS coherence length in the bulk
superconductor, and we can use the previous solution with the
step-like form for $\Delta$. The pair potential is given by the
self-consistency equation
\begin{equation}
\label{self} \Delta(z)= 2\lambda N(0)\int d^2{\bf
k}_{||}\int_0^{\hbar\omega_D}d\Omega~u({\bf r})v^\ast({\bf
r})\tanh(E/2k_BT),
\end{equation}
where $\lambda$ is the BCS coupling constant and
$N(0)=mk_F/2\pi^2\hbar^2$ is the normal-metal density of states
(per spin) at the Fermi level. We calculate the spatial average of
$\Delta(z)$ using the standard iteration procedure
\begin{equation}
\Delta_{i+1}= \frac{1}{d}\int_0^d\Delta_i(z)dz,
\end{equation}
by setting $\Omega=\sqrt{E^2-\Delta^2_i}$ in Eq. (\ref{self}) in
order to obtain $\Delta_i(z)$ in the $i$-th iteration.\cite{slatt}
Starting from the bulk value $\Delta_0$, we repeat this procedure
until the difference between $\Delta_{i+1}$ and $\Delta_i$ becomes
sufficiently small. In Fig. \ref{delta}, we show the resulting
$\Delta$ as a function of $d$, at zero temperature.


\section{Differential conductance and density of states}

The normalized differential conductance at zero temperature for a
planar double junction is
\begin{equation}
\label{conduct}
\frac{G(E)}{G_N}=\int_0^{\pi/2}d\theta~\sin\theta\cos\theta~\left[1+A(E,\theta)\right],
\end{equation}
where $G_N=e^2/h$. In the one-dimensional (1D) case, the
normalized differential conductance is simply
\begin{equation}
\label{1Dconduct} \frac{g(E)}{G_N}=1+A(E,0).
\end{equation}

Coherent transport through an NSN double junction is influenced by
considerable changes of the quasiparticle density of states with
respect to the bulk superconductor.\cite{Anderson,Rowell} Adopting
the method introduced by McMillan,\cite{McMillan}
Ishii,\cite{Ishii} and Furusaki and Tsukada,\cite{Furusaki
Tsukada} we evaluate the Green functions by combining the
solutions of the Bogoliubov-de Gennes equation. From the imaginary
part of the retarded Green function, we obtain the local value of
the partial density of states (PDOS) for the superconducting film,
normalized with respect to the 1D normal-metal density of states
per spin at the Fermi level,
\begin{eqnarray}
\tilde{N}(z,\theta,E)=\frac{1}{\Gamma(E)\cos\theta}& {\rm Re}&
\Big\lbrace 2E^2(E^2+\Omega^2)+2E^2\Delta^2\cos\zeta \nonumber\\
&~&+[\cos(\zeta(z/d-1))+\cos(\zeta z/d)]
[\Delta^4-\Delta^2(E^2+\Omega^2)\cos\zeta]\nonumber\\ &~&+
2E^2\Delta^2[\sin(\zeta(z/d-1))-\sin(\zeta
z/d)]\sin\zeta\Big\rbrace~,
\end{eqnarray}
where
\begin{equation}
\Gamma(E)=[(E^2+\Omega^2)
\cos\zeta-\Delta^2]^2+4E^2\Omega^2\sin^2\zeta.
\end{equation}
For a planar junction, the local DOS for the superconducting film
is given by
\begin{equation}
\frac{N(z,E)}{N(0)}=\int_0^{\pi/2}d\theta\sin\theta\cos\theta~\tilde{N}(z,\theta,E).
\end{equation}
Here, $\tilde{N}(z,\theta,E)=N(z,E)/N(0)=1$ in the normal-metal
electrodes ($z<0$ and $z>d$).

Characteristic features of the coherence in NSN double junctions
are subgap transport of electrons and oscillations of the
conductance. Due to the interference effect, the Andreev
reflection is suppressed for $E<\Delta$, whereas the AR
probability oscillates with $E$ and $d$, for $E>\Delta$. The
subgap transport is dominant in thin superconducting films, and
the oscillatory behavior is apparent in thick films. These
oscillations are more pronounced in the 1D case, which is shown in
Fig. \ref{E2n2}. It can be clearly seen that the positions of the
minima in $g(E)$ exactly match the positions of the maxima in
PDOS, corresponding to the geometrical resonances imposed by Eq.
(\ref{resonance}). Results can be applied to spectroscopic
measurements of $\Delta$ and $v_F$. From Eq. (\ref{resonance}) for
$\theta=0$, the energies of the geometrical resonances, $E_n$,
satisfy
\begin{eqnarray}
\label{En} E_n^2=\Delta^2+\Big(\frac{\pi\hbar v_F}{d}\Big)^2n^2,
\end{eqnarray}
where $n=1,2,\ldots$ counts the conductance minima. Therefore, the
plot of $E_n^2$ vs $n^2$ has the intercept equal to $\Delta^2$,
and the slope equal to $\tan^{-1}\big[(\pi\hbar v_F/d)^2\big]$. An
example is shown in the Inset of Fig. \ref{E2n2}. Note that the
points obtained from the minima of $G(E)$ follow practically the
same linear law as those of $g(E)$.

The obtained results are in a strong contrast with the conductance
spectrum and DOS of tunnel junctions. In the latter case, $g(E)$
and PDOS reduce to $\delta$-function-like spikes at the bound
state energies of the isolated superconducting film, given by
$lq^+=n_1\pi$ and $lq^-=n_2\pi$, where $n_1-n_2=2n$, with $n$
given in Eq. (\ref{resonance}).\footnote{See Fig. 2 in M. Bo\v
zovi\'c and Z. Radovi\'c, cond-mat/0207375 (2002).}

The conductance spectra $G(E)/G_N$ and the local DOS for the
superconducting film in a planar NSN double junction are
illustrated in Figs. \ref{G} and \ref{DOS}, for $d/\xi_0=1$ and
$d/\xi_0=10$, using the self-consistent pair potential
$\Delta/\Delta_0=0.816$ and $\Delta/\Delta_0=1$, respectively. In
Fig.~\ref{G}, the conductance spectra are compared to the BTK
result describing the incoherent transport. It can be seen in
Fig.~\ref{DOS}(a) that the energy spectrum of a thin
superconducting film is practically gapless. For a thick
superconducting film, this is the case only near the interfaces,
Fig.~\ref{DOS}(b). The characteristic oscillations of the local
DOS become more apparent for a thick film. The oscillations of
$G(E)$ and of the local DOS are not completely smeared in the 3D
case. The positions of the minima in $G(E)$ or the maxima of the
local DOS are still close to the positions of the resonant
quasiparticle states in the 1D case.


\section{Conclusion}

We have derived simple expressions describing the finite size and
coherency effects in the metallic NSN double junctions, suitable
for experimental data analysis. The main features of the coherent
quantum transport through a superconducting film are subgap
transport and oscillations of the differential conductance. The
Andreev reflection is suppressed below the gap, especially in thin
films. The conductance oscillates as a function of the layer
thickness and of the quasiparticle energy above the
superconducting gap. Periodic vanishing of the Andreev reflection
at the energies of geometrical resonances is an important
consequence of the quasiparticle interference; the conductance
minima correspond to the maxima in the density of states.
Measurements of the conductance spectra of metallic NSN double
junctions could be used for accurate determination of the pair
potential and the Fermi velocity in superconductors.

\bigskip

We are grateful to Ivan Bo\v zovi\'c for useful discussions. This
work has been supported by the Serbian Ministry of Science,
Technology, and Development, Project ${\rm N}^{\circ} 1899$.


\begin{figure}[hb]
\caption { \label{delta} The pair potential $\Delta$ as a function
of the film thickness $d$, at zero temperature. }

\caption{\label{E2n2} (a) Partial ($\theta=0$) conductance
spectra, $g(E)/G_N$, and (b) corresponding PDOS,
$\tilde{N}(d/2,0,E)$, for $d/\xi_0=10$. Inset: square of resonant
energies, $E^2_n$, vs $n^2$ obtained from the minima of $g(E)$
(open circles) and $G(E)$ (full squares).}

\caption{\label{G} Differential conductance spectra $G(E)/G_N$ for
$d/\xi_0=1$ and $10$ (solid curves). The BTK result is shown for
comparison (dashed curve). }

\caption{\label{DOS} Local DOS, $N(z,E)/N(0)$, for $z=d/2$ (solid
curves) and $z=0$ or $d$ (dotted curves) as a function of
$E/\Delta_0$ is shown for $d/\xi_0=1$ (a), and $d/\xi_0=10$ (b).}
\end{figure}

\pagebreak

\begin{figure}[h]
\dimen255=0.5\textwidth
    \centerline{\psfig{figure=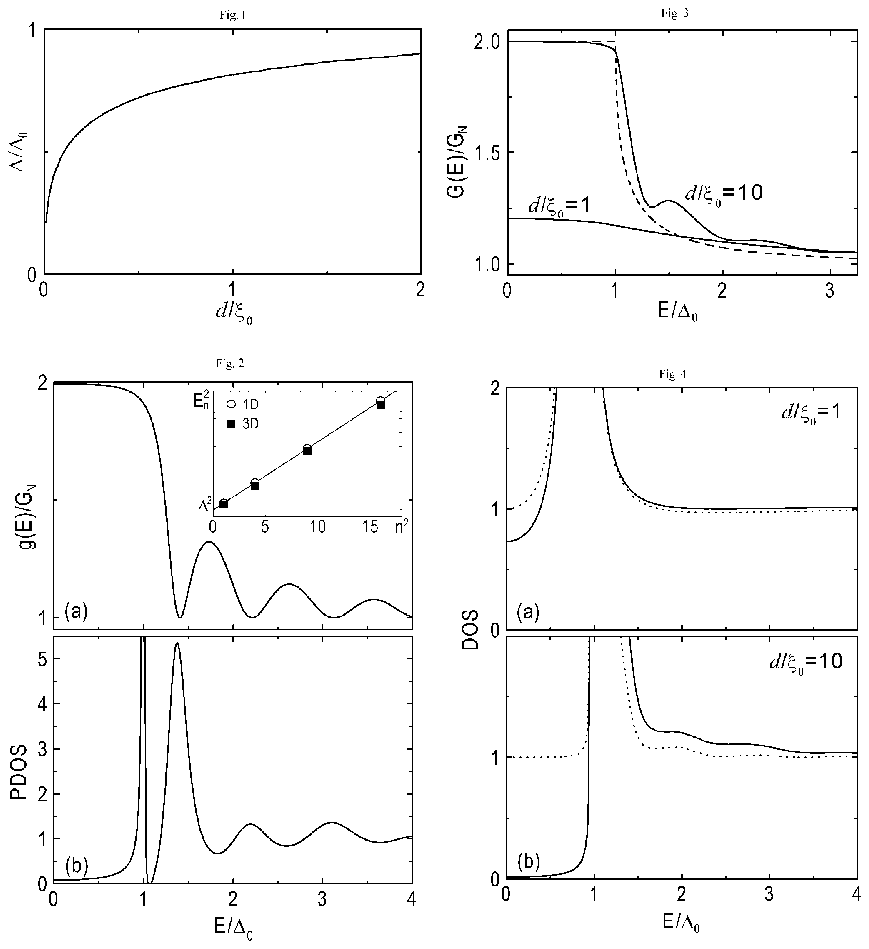,height=2\dimen255}}
\end{figure}

\end{document}